# Microstructure and corrosion evolution of additively manufactured aluminium alloy AA7075 as a function of ageing

O. Gharbi[1,2,+], S.K. Kairy[2,+], P.R. De Lima[2], D. Jiang[2], J. Nicklaus[2] and N. Birbilis[3]

[1]Laboratoire interfaces et systèmes électrochimiques, CNRS, Sorbonne Université UMR8235, Paris, France.
[2]Department of Materials Science and Engineering, Monash University, Clayton, VIC 3168, Australia.
[3]College of Engineering and Computer Science, The Australian National University, Acton, ACT 2601, Australia.

[1]oumaima.gharbi@sorbonne-universite.fr
[+] These authors contributed equally

**Abstract**

Additively manufactured high strength aluminium (Al) alloy AA7075 was prepared using selective laser melting (SLM). High strength Al-alloys prepared by SLM have not been widely studied to date. The evolution of microstructure and hardness, with the attendant corrosion, were investigated. Additively manufactured AA7075 was investigated both in the "as-produced" condition and as a function of artificial ageing. The microstructure of specimens prepared was studied using electron microscopy. Production of AA7075 by SLM generated a unique microstructure, which was altered by solutionising and further altered by artificial ageing – resulting in microstructures distinctive to that of wrought AA7075-T6. The electrochemical response of additively manufactured AA7075 was dependent on processing history, and unique to wrought AA7075-T6, whereby dissolution rates were generally lower for additively manufactured AA7075. Furthermore, immersion exposure testing followed by microscopy, indicated different corrosion morphology for additively manufactured AA7075, whereby resultant pit size was notably smaller, in contrast to wrought AA7075-T6.

# 1. Introduction

Metal additive manufacturing (AM) is a promising technology for complex and net-shape prototyping and production, with minimal waste generation [1-5]. With the prospect of net shape production of high strength aluminium (Al) alloys [6-12], the interest in AM technology to produce components from high strength 7xxx series (Al-Zn-Mg(-Cu)) Al-alloys is increasing [6, 13-18]. To date, the microstructural evolution and corrosion behaviour of AM prepared 7xxx series Al-alloys have not been studied and hence not understood in contrast to wrought 7xxx series Al-alloys presently used in aerospace applications.

The microstructure of wrought aluminium alloy AA7075, which is conventionally produced from homogenised cast ingots by either rolling, extrusion or forging, has been extensively studied [19-28]. Wrought AA7075-T6, as a plate or sheet product, principally contains nanometre scale η′-phase (which nominally $MgZn_2$) precipitates as the major strengthening phase. However, several second phase particles (including dispersoid phases or non-strengthening constituent particles), such as $Al_7Cu_2Fe$, $Al_3Fe$, $Mg_2Si$, $Al_2Cu$, $Al_2CuMg$, $MgZn_2$, $Al_6Mn$, $Al_3Ti$ and $Al_3Zr$, with varying size, shape and distribution have been previously determined to populate wrought AA7075 [19-28]. Second phase particles influence corrosion behaviour of wrought AA7075 [26, 29-35].

The localised corrosion, and its morphology, of wrought 7xxx Al-alloys is principally associated with second phase particles, as the particles exhibit electrochemical characteristics different from those of the Al-matrix [26, 29-45]. Electrochemical characteristics of several second phase particles that populate in the Al-matrix of wrought AA7075 have been previously determined in solutions of various chloride concentrations and pH, either by the production of bulk intermetallic phases, by the electrochemical microcell technique, or by electron microscopy [30, 38, 44-52] – providing a deterministic rationale for the of each second phases in the localised corrosion of AA7075. It was determined that $Al_7Cu_2Fe$, $Al_3Fe$, $Al_6Mn$, $Al_3Ti$, $Al_3Zr$ and $Al_2Cu$ are 'cathodic' to the Al-matrix; whilst, $Mg_2Si$, $Al_2CuMg$ and $MgZn_2$ are

'anodic' to the Al-matrix. At sites containing cathodic particles, localised corrosion develops in the form of circumferential trenches by dissolving the Al-matrix at the particles' periphery [26, 30-31, 44-45]. Whilst at anodic sites, 'self-dissolution' of the second phase particle occurs, most typically in the form of incongruent particle dissolution / dealloying of Mg from such anodic second phase particles ($Mg_2Si$, $Al_2CuMg$ and $MgZn_2$) resulting in remnants of either Si-, $Al_2Cu$- and Zn-, respectively [38, 46-52]. As has been documented previously, following its dealloying, the remnant of $Al_2CuMg$ has been determined to behave as a local cathode, dissolving the Al-matrix [50-52], the role of $Mg_2Si$ and $MgZn_2$ (the latter of which may be Cu enriched [42]) remnants is not clear. Pitting and localised corrosion associated with second phase particles evolves with time, and serves as the precursor to intergranular corrosion in wrought AA7075-T6, which is an alloy that can also suffer stress corrosion cracking in the presence of tensile stress [31, 33-34, 35-41]. Further, pits in AA7075-T6 also well documented to act as initiation sites for the nucleation of fatigue cracks as a result of cyclic loading [53-55].

To date, studies on the microstructure and mechanical properties of SLMed 7xxx Al-alloys are limited [6, 13-18], and in the context of corrosion, no prior studies exist to date. A recent study discovered the predominant presence of a nanometre scale icosahedral quasicrystalline particle, termed ν-phase [13]. This (ν) phase was comprised of Zn-Mg-Cu(-Al), a structurally and chemically unique in the Al-matrix of as-SLMed AA7075 [13]. As a consequence of this discovery, a more detailed study upon as-SLMed AA7075, including as a function of artificial ageing, is essential in the context of understanding microstructural evolution – and also the corrosion response in contrast to wrought AA7075-T6.

Herein, the microstructure, hardness, electrochemical behaviour and localised corrosion response of SLMed AA7075 in three different conditions, viz. as-SLMed, solutionised + quenched, and solutionised + quenched and aged at 120°C for 24 h, were studied and compared with those of wrought AA7075-T6.

## 2. Experimental

### 2.1 Materials and ageing

Additively manufactured AA7075 specimens were produced from AA7075 powder using selective laser melting (SLM). The AA7075 powder, with size distribution of 20-63 µm, was supplied by LPW technology. A Concept Laser Mlab cusing-R was used for SLM, described in detail in [13], with the production conditions for specimens tested herein consisting of; laser power, 95 W; laser scan speed, 200 mm/s; and layer thickness, 25 µm. In addition to SLM prepared specimens, wrought AA7075-T6 rod was also studied for comparison. The compositions of the AA7075 powder as supplied, as-SLMed AA7075 and wrought AA7075-T6 were determined by inductively coupled plasma - optical emission spectroscopy (ICP-OES) at an accredited laboratory (Spectrometer services, VIC. Australia) and shown in Table 1.

As-SLMed AA7075 studied herein presented a consolidation of > 98%. In order to study microstructural evolution, hardness and corrosion, as-SLMed AA7075 specimens were solutionised at 480°C for 1.5 h in a salt bath and quenched in water; followed by artificial aging at 120°C (for different times up to 110 h) in an oil bath, followed by water quenching.

### 2.2 Microstructural characterisation

#### 2.2.1 Scanning electron microscopy

Scanning electron microscopy (SEM) was performed using a JEOL® 7001F SEM operated at 15 kV. Energy-dispersive X-ray spectroscopy (EDXS) was performed using an Oxford Instruments® X-Max 80 silicon detector, and analysed via Aztec® software. Specimens for SEM were metallographically prepared to a 0.01 µm surface finish under ethanol and colloidal silica suspension.

#### 2.2.2 Transmission electron microscopy

Scanning transmission electron microscopy (STEM) was performed using an FEI® Tecnai $G^2$ F20 S-TWIN FEG operated at 200 kV and an aberration corrected FEI® Titan$^3$ 80-

300 operated at 300 kV. Imaging was performed using a high-angle annular dark-field (HAADF) detector, which provides atomic number contrast [57]. To determine the composition of different phases, EDXS was performed using a Bruker® X-flash X-ray detector (Tecnai G$^2$ F20), and the EDXS data was analysed using Bruker® Esprit X-ray software.

Specimen preparation for TEM involved punching 3 mm discs from thin slices of SLMed AA7075 specimens followed by mechanical grinding under ethanol to a thickness of ~ 60 µm, using 1200 grit SiC paper, and then argon ion milling at -50°C using a Gatan® precision ion polishing system (PIPS). Specimens were plasma cleaned prior to imaging for ~ 2 min with an argon and oxygen gas mixture using a Gatan® solarus 950 advanced plasma system.

## 2.3 Hardness testing

Vickers hardness testing was conducted by applying 1 kg of load for 10 s using a Duramin A-300 hardness tester. Specimens for hardness testing were prepared to an 800 grit SiC paper finish, and five measurements were collected for each specimen.

## 2.4 Electrochemical characterisation

Potentiodynamic polarisation testing was performed using a Bio-logic VMP-3, using a 3-electrode flat cell (inclusive of a saturated calomel reference electrode and platinum mesh counter electrode). Specimens for polarisation testing were prepared to a 2000 grit SiC paper finish under ethanol, and ultrasonically cleaned in ethanol. Testing was carried out in 0.1 M NaCl. Prior to polarisation testing, specimens were exposed at OCP for 30 mins. Separate anodic and cathodic scans were conducted at 1 mV.s$^{-1}$ scan rate by commencing at ± 20 mV vs. OCP and ceasing at absolute current of 10 mA.cm$^{-2}$. Each scan was reproduced at least in triplicate for reproducibility.

Electrochemical impedance spectroscopy (EIS) was performed in naturally aerated 0.1 M NaCl using a frequency range of 10 kHz to 10 mHz (with testing at 10 points per decade) and an applied potential of +/- 10 mV. Each test was replicated at least in triplicates for reproducibility.

# 3. Results and discussion

## 3.1 Microstructural analysis

### *3.1.1 Scanning electron microscopy analysis*

The microstructures of AA7075-T6 and SLMed AA7075 in three different conditions, viz. as-SLMed, solutionised + quenched, and solutionised + quenched and aged at 120°C for 24 h, were realised using SEM in backscattered electron (BSE) mode (Figure 1).

Specimens from wrought AA7075-T6 exhibited uniformly distributed coarse (up to ~15 µm in size) second phase particles in the Al-matrix (Figure 1a), with such coarse second phase particles often forming networks that may extend up to ~100 µm in size (Figure 1a, inset). The coarse particles were identified as Al-Cu-Fe(-Si) phase constituents, as determined from EDXS point analysis, which are nominally formed during the casting process [19] and do not adequately redissolve during subsequent thermomechanical processing. In contrast, as-SLMed AA7075 revealed a characteristic melt pool structure (synonymous with SLM prepared alloys) with high density nanometre to sub-micrometre scale Mg-Zn-Cu(-Al) and $Mg_2Si$ (β-phase) second phase particles. The Mg-Zn-Cu(-Al) and $Mg_2Si$ (β-phase) particles were homogenously distributed in the Al-matrix (Figure 1b), emphasising the impact of high solidification rates (> $10^6$ °C/s [3, 7, 16, 18]) that are achieved during the SLM process. The second phase particles in as-SLMed AA7075 were mostly spherical; however, some elongated and thread-like shaped particles were also observed along melt pool boundaries (Figure 1b). In addition, the second phase particles were considerably finer within the core of melt pools and coarser at melt pool boundaries. Such difference in particle size, shape and distribution at melt pool boundaries when compared to those within the core of melt pools may be rationalised by the cooling rate gradient within each melt pool. Melt pool boundaries experience lower cooling rates than the core of melt pools [5, 7, 16-17, 58], and thus may resulting in greater solute rearrangement at the boundaries (although noting that the attendant cooling rates are likely to significantly inhibit diffusion).

It is noted that the SLM process itself, led to loss of Zn and Mg, whilst Cu, Cr and Fe were not lost, as determined from ICP-OES compositional analysis of the AA7075 feedstock powder and as-SLMed AA7075 (Table 1). In such high-power laser based AM processing, elements with low fusion are susceptible to vaporising as a result of their higher equilibrium vapour pressure than that of Al [59]. Loss of alloying elements as vapour was determined to increase scan track instability, porosity and solidification cracking in AM specimens [59]. It is also noted that the SLM process annihilated coarse Fe-containing constituent particles in as-SLMed AA7075. The annihilation of coarse Fe-containing constituent particles in as-SLMed AA7075 is beneficial in the context of localised corrosion susceptibility, since localised corrosion predominantly associated with such particles in wrought AA7075-T6 [26, 29-45]. In addition, the loss of Zn and Mg may also influence the artificial aging of SLMed AA7075, as Zn and Mg are essential for forming $MgZn_2$ phase precipitates in the Al-matrix.

The effect of solutionising on the microstructure of SLMed AA7075 can be observed from Figure 1c. The majority of the sub-micrometre sized particles as well as the melt pool boundaries observed in the as-SLMed condition are no longer perceptible in the solutionised condition (at identical scales). At the SEM image scale, it is corroborated the dissolution of the majority of particles in the Al-matrix of SLMed AA7075 resulted in super saturated solid solution (Figure 1c). The undissolved second phase particles, as shown in the inset in Figure 1c, were determined to be Al-Cu-Fe(-Si) particles, which do not dissolve upon solutionising [19]. However, the size and number density of such coarse Fe-containing constituents in solutionised SLMed AA7075 was significantly lower than in wrought AA7075-T6. Further ageing at 120°C for 24 h resulted in the nucleation and growth of precipitates along grain boundaries and within grains (Figure 1d). As the diffusion of solute is higher along grain boundaries than within grains [60], precipitates were larger along grain boundaries, revealing remnant columnar grains in the microstructure (Figure 1d). From EDXS-SEM point analysis, it was identified that the second phase particles in aged SLMed AA7075 are Al-Cu-Fe(-Si) constituent particles and precipitates containing Zn, Mg and Cu (η-phase). Very fine pits (dark in contrast) corresponding to second phase particles in Figure 1(a-d) were formed as a result of final polishing in colloidal silica suspension.

The SEM analysis overall revealed significant microstructural variation between wrought AA7075-T6 and SLMed AA7075. In addition, the microstructure of as-SLMed AA7075 also varied significantly with solutionising and artificial ageing. Such microstructural variations are important to understand as they are significant in influencing the hardness, electrochemical behaviour and localised corrosion and of AA7075 produced by SLM.

### 3.1.2 Transmission electron microscopy analysis

Detailed TEM characterisation of SLMed AA7075 in three different conditions, viz. as-SLMed, solutionised + quenched, and solutionised + quenched and aged at 120°C for 24 h, was performed and the resulting images are shown in Figures 2 to 4.

Second phase particles with sizes less than 200 nm were observed in SLMed AA7075 (Figure 2 (a-c)). Another significant microstructural feature, cellular structures were observed in SLMed7075, as pointed out by solid arrows in Figure 2(a′-c′). The size of such cellular structure was found to increase following solutionising. Cellular structurers were realised by the segregation of high atomic number elements, most probably Zn and Cu, along the structure boundaries, as evident from the HAADF images. Following ageing at 120°C for 24 h, nanoscale η′-phase ($MgZn_2$) precipitates evolved in the Al-matrix, mostly along cellular structure boundaries, as shown by the dotted arrow in Figure 2c′, indicating that the solute segregated along the cellular structure boundaries aid in nucleating η′-phase precipitates via solute diffusion and redistribution. In fact, η′-phase precipitates were also found to contain Cu, as previously reported in wrought Cu-containing 7xxx Al-alloys [38, 42, 61].

Second phase particles in as-SLMed AA7075 were often found to co-exist together (rather vividly shown in Figure 3 a and b), indicating that the first nucleated phases were likely to provide sites for the heterogeneous nucleation of other phases. The mechanism of such nucleation processes is complex, considering high cooling rates and subsequently negligible solute diffusion rates during the SLM process [1-7, 59]. However, most of the particles in as-SLMed AA7075 were characterised to be of icosahedral quasicrystalline phase (Figure 3c), termed as ν-phase (Mg-Zn-Cu(-Al)) [13]. As a result of coexistence of different phase particles at a single position, the determination of accurate composition of such particles is challenging. However, an estimation of phases based on the compositional analysis via the EDXS-STEM technique can be realised (Table 2). It is noted that dissolution of quasicrystalline ν-phase following solutionising was discovered, implying that ν-phase is a metastable phase, most likely only existing in as-SLMed AA7075. Such important characteristics of quasicrystalline ν-phase were not previously determined, nor the relevance in the context of designing SLMed AA7xxx alloys.

A nanoscale platelet-shaped phase (< 1 nm in thickness) that is considered to be the precursor to η′-phase was formed on the $\{111\}_\alpha$ planes of the Al-matrix in as-SLMed AA7075, as determined from the FFT analysis and atomic resolution image (Figure 4a). The precursor η′-phase could have formed either during rapid solidification achieved via the SLM process or potentially during localised reheating of an already solidified layer during the SLM process. However, the phase was less in number and not uniformly distributed in the Al-matrix. Precursor η′-phase was not observed following solutionising (Figure 4b), indicating the dissolution of the phase. Nanoscale η′-phase precipitates evolved in the Al-matrix following ageing at 120°C (Figure 4c), caused by solute diffusion and redistribution. It is noted that η′-phase precipitates, which evolved during the ageing process, were not uniformly distributed in the Al-matrix. In contrast, such precipitates evolve uniformly in the Al-matrix of wrought

AA7075-T6 [19-20, 23, 25, 27-28, 61]. The orientation relationship of the η′-phase precipitate with the Al-matrix of 24 h aged SLMed AA7075 was identified, and was similar to that for wrought AA7xxx alloys [19-20]: $(0001)_{η′}$ // $\{111\}_α$ , $<11\bar{2}0>_{η′}$ // $<11\bar{2}>_α$.

### 3.2 Hardness studies

The variation in Vickers hardness of SLMed AA7075 as a function of ageing is presented in Figure 5.

The average hardness of as-SLMed AA7075 decreased from ~ 107 to 64 VHN following solutionising. Ageing at 120°C caused the hardness to gradually increase with time, and following 24 h the hardness reached to that of the value recorded for as-SLMed specimens. The hardness further increased with time and reached a maximum value of ~ 126 VHN following 110 h of ageing. In contrast, the average Vickers hardness of wrought AA7075-T6 was determined to be 197 ± 5 VHN. The variation in hardness between SLMed AA7075 and wrought AA7075-T6, is a result of the corresponding microstructural variation, but also it is noted that the bulk composition of SLMed AA7075 is slightly alloy lean by comparison, and that SLMed AA7075 has not undergoing any cold (or hot) working. The second phase particles, nanoscale precipitates and cellular structures in the Al-matrix of SLMed AA7075 as shown in Figures 1-4 and Table 2, contributed to hardness. The decrease in hardness following solutionising is mainly attributed to the predominant dissolution of quasicrystalline ν-phase and nanometre scale precursors of η′-phase, along with the growth of cellular structures. The evolution of η′-phase precipitates in the Al-matrix following ageing at 120°C resulted in an increase in hardness. However, the number density and distribution of η′-phase precipitates in artificially aged SLMed AA7075 were significantly lower and non-uniform, respectively, than those reported in wrought AA7075-T6 alloys [19-20, 23, 25, 27-28, 61].

**3.3 Electrochemical analysis**

*3.3.1 Potentiodynamic polarisation*

Separate anodic and cathodic potentiodynamic scans of SLMed AA7075 and wrought AA7075-T6 were conducted, and the results are presented in Figure 6.

The anodic and cathodic polarisation response of SLMed AA7075 indicate a shift in the corrosion potential ($E_{corr}$) from - 0.9 to - 1.3 $V_{SCE}$ following solutionising (where the maximum level of Mg and Zn is in solid solution) and then to -1.2 $V_{SCE}$ following 24 h ageing. In addition, in the case of as-SLMed AA7075, the anodic current significantly increased at potentials above $E_{corr}$. Following solutionising, a distinct shift between $E_{corr}$ and $E_{pit}$ (a potential above which stable pitting can occur) is notable. In fact, this plateau of approximately 400 mV (between $E_{corr}$ and -0.7 $V_{SCE}$) correspond closely to the $E_{corr}$ of wrought AA7075-T6. The presence of such significant shift of $E_{corr}$ following solutionising and ageing coincides with the dissolution of quasicrystalline ν-phase particles in the Al matrix. More importantly, $E_{pit}$ of SLMed AA7075 following solutionising and ageing remains similar to the $E_{corr}$ of wrought AA7075-T6, suggesting a similar electrochemical response – which corroborates with a scenario of the closest of the microstructures between the SLMed AA7075 and the wrought AA7075-T6.

The evolution of the cathodic polarisation section did not reveal a significant difference in cathodic kinetics between as-SLMed and solutionised AA7075. In addition, the cathodic kinetics are slightly decreased (from $\sim 10^{-2}$ to $\sim 5.10^{-3}$ mA.cm$^{-2}$) between $E_{corr}$ to -1.4 $V_{SCE}$ between wrought AA7075-T6 and SLMed AA7075. This may be attributed to the refinement and annihilation of coarse 'cathodic' constituent particles in the SLMed AA7075. Similar results were reported previously, where a minor change in the cathodic kinetics was observed for an additively manufactured AA2024 [9]. However, in the case of additively manufactured AA2024 although the overall electrochemical kinetics were shown to be similar for the wrought and SLMed conditions (noting that AA2024 is Cu-rich alloy dominated by cathodic particles), the refined microstructure obtained *via* SLM positively influenced the localised corrosion susceptibility of the AA2024 with a decrease of the local cathode: anode ratio, limiting the onset of local pH increase (from oxygen reduction reaction) and Al matrix dissolution.

Overall, the variation in electrochemical response of SLMed AA7075 with thermal treatment, and also compared to that of wrought AA7075-T6, is a result of significant microstructural variation and corresponding compositional variation in the Al-matrix.

### *3.3.2 Electrochemical impedance spectroscopy*

The EIS results of wrought AA7075-T6 and SLMed AA7075 are presented as Nyquist and Bode plots in Figure 7 (a-c). Equivalent electrical circuits to fit experimental EIS data of SLMed AA7075 and wrought AA7075-T6 are presented in Figure 7 d and e, respectively. The associated parameters obtained via fitting of EIS data were determined and presented in Table 3. In addition, the resistance ($R_4$) and inductance ($L_1$) corresponding to the inductive loop of wrought AA7075-T6 were also determined (Table 3). In the tabulated data, the polarisation resistance ($R_P$) corresponds to the extrapolated intersection of the impedance data with $-Z'' = 0$ at very low frequencies (in Figure 7a).

The EIS spectrum of wrought AA7075-T6 characteristically displayed two capacitive loops at high or moderate frequencies and an inductive loop at low frequencies (Figure 7a). The inductive behaviour for wrought AA7075-T6 can also be observed in the Bode magnitude plot (Figure 7c) at low frequencies ($f < 0.2$ Hz). Wrought 7xxx Al-alloys in NaCl were previously determined to exhibit inductive loop at low frequency regime in the Nyquist plot [62-63]. The presence of such inductive loop indicates a degradation of passive or oxide film developed upon Al alloys [62-63], and corrosion during the duration of testing. In addition, wrought AA7075-T6 freely corrodes at open circuit conditions in NaCl and therefore, the passive film over the alloy is not sustained. In contrast to wrought AA7075-T6, SLMed AA7075 did not exhibit a characteristic inductive loop, indicating that the film developed upon SLMed AA7075 is expected to provide corrosion protection comparatively greater than for wrought AA7075-T6 in NaCl. In addition, it was determined that $R_p$ was greater for SLMed AA7075 compared to wrought AA7075-T6. The $R_p$ of as-SLMed AA7075 was also found to vary with solutionising and ageing.

**3.4 Localised corrosion response**

To study the localised corrosion response, specimens of wrought AA7075-T6 and SLMed AA7075 were immersed in 0.1 M NaCl for 10 h and subsequently examined, both prior to and following immersion using SEM (Figures 8-11).

Upon wrought AA7075-T6, localised corrosion was most commonly associated with Al-Cu-Fe(-Si) phase constituent particles (Figure 8 b-d). Pitting can be identified as dark the regions in Figure 8b. Upon closer observations (Figure 8 c-d), pits were developed in the form of circumferential trenches by dissolving the Al-matrix at the periphery of Al-Cu-Fe(-Si) phase constituent particles. The results obtained herein support the documented corrosion morphology of high strength Al-alloys with respect to the role of Fe-containing particles serving as local cathodes (and supporting oxygen reduction at enhanced rates relative to the matrix) and leading to trenching in NaCl solutions [29-31, 36, 44-45]. In addition, enhanced oxygen reduction at local sites will generate a hydroxyl ion atmosphere at cathodic particles, thereby increasing the local pH and accelerating the dissolution of the Al-matrix at the periphery of such particles [31]. A few, and small, pits that were not associated with Al-Cu-Fe-(Si) phase constituent particles were also observed in the Al-matrix (Figure 8b). Such pits may have however formed at the periphery of Al-Cu-Fe(-Si) particles that would have detached following a period of Al-matrix dissolution at the particles' periphery or could be dealloying of active phases such as $Mg_2Si$ and Cu-containing $MgZn_2$, which were very fine and could not be detected under SEM.

Specimens from as-SLMed AA7075 also exhibited pitting that was associated with Mg-Zn-Cu(-Al) (ν-phase) particles in the Al-matrix (Figure 9 b-d). Pits were developed in the form of trenches by dissolving the Al-matrix at the periphery of ν-phase (Figure 9 b-d), revealing the apparently cathodic nature of quasicrystalline ν-phase particles, whose electrochemical behaviour was previously not known. In addition, the constituent $Mg_2Si$ particles suffered dealloying by Mg dissolution, which is consistent with previous findings [46-48]. Pitting was more pronounced along the melt pool boundaries than that within the core of melt pools (Figure 9 b-c). Such variation in pitting behaviour along melt pool boundaries and within the core of melt pools is a result of the microstructural variation caused by SLM production; particles along the boundaries were larger and in a lower number density compared to those within the core of melt pools.

Solutionising as-SLMed AA7075 specimens caused the subsequent pitting morphology to evolve uniformly in the Al-matrix (Figure 10b). In addition, the pit number density decreased in the solutionised condition when compared to the as-SLMed condition. The decrease in pit number density and distribution following solutionising is a direct result of the dissolution of ν-phase and disappearance of melt pools. Pits in the solutionised condition were associated with dissolution of the Al-matrix at the periphery of Al-Cu-Fe-(Si) particles (Figure 10 b-d). The overall dissolution of the Al-matrix also increased following solutionising, as evident from cracks developed upon the oxide layer several micrometres in thickness (Figure 10b), validating the potentiodynamic polarisation testing results in a general sense. The increased dissolution of the Al-matrix following solutionising may be attributed to the composition of the Al-matrix, which is a supersaturated solid solution of Al principally containing the Zn and Mg in solid solution, which was evidenced to also decrease the $E_{corr}$ rather vividly in the solutionised condition.

In response to immersion for 10h in 0.1M NaCl, SLMed AA7075 specimens aged at 120°C for 24 h revealed the formation of pits with greater size along grain boundaries than within grains (Figure 11b). This is associated with larger second phase particles and precipitates along grain boundaries, than within grains. The pits observed associated with second phase particles were also in the form of trenches, dissolving the Al-matrix at the periphery of cathodic Al-Cu-Fe(-Si) phase constituent particles (Figure 11 c-d). Conversely, the η′-phase ($MgZn_2$) precipitates containing Cu, which formed along grain boundaries in response to artificial aging, are considered to have undergone dealloying by selective Mg-dissolution, however this was unable to be resolved using SEM.

The statistics related to pitting as determined from numerous SEM images (including Figures 8-11) of post corrosion surfaces from wrought AA7075-T6 and SLMed AA7075 are presented in Table 4.

Post immersion SEM (Figures 8-11) and the results in Table 4, it may be rationalised that pits developed upon wrought AA7075-T6 and SLMed AA7075 were associated with second phase particles. In addition, the size, shape and distribution of second phase particles influenced pit size, shape and distribution – indicating that localised corrosion is deterministic. In comparison with wrought AA7075-T6, pits developed upon SLMed AA7075 were greater in number, however notably smaller in size as a result of the increased population and decreased size of second phase particles present in the Al-matrix of specimens generated by SLM. In addition, the contribution of 'cathodic corrosion' as the mechanism for generating trenching at the periphery of second phase particles in SLMed AA7075 is also expected to be minimal, as nanometre to sub-micrometre scale cathodic second phase particles do not substantially increase the local pH when compared to the same particles in micrometre size.

### 3.5 General discussion

The localised corrosion caused by pitting, even from the early stages of corrosion in wrought AA7075-T6, can initiate fatigue cracks in the presence of cyclic stress - leading to the failure of components prior to the service life [53-55]. Therefore, it is relevant to discuss the role of pitting developed upon SLMed AA7075 in contrast to the localised corrosion of wrought

AA7075-T6. It was previously reported that, pits with depths of 29-75 µm initiated fatigue cracks in wrought AA7075-T6 at a nominal maximum tensile stress of ~210 MPa [36, 54]. In addition, the total fatigue life decreased with the severity of corrosion damage [54-55, 64], although the number of cycles to initiate a fatigue crack did not depend on the depth of pits [65-66]. Herein, it was identified that pits developed upon wrought AA7075-T6 were larger, and expected to be much deeper, by several orders of magnitude when compared to those upon SLMed AA7075. Therefore, it can be understood that the contribution of such fine pits on fatigue corrosion of SLMed AA7075 may be less significant, and therefore an important avenue for future work. Moreover, the impedance results herein revealed the persistence of a comparatively higher corrosion resistance for SLMed AA7075. However, the results herein indicate that future systematic studies regarding the damage tolerance of SLMed AA7075 are required, which should also include the testing of stress corrosion cracking susceptibility. The results herein however, represent a first presentation of the corrosion of SLMed AA7075, including in response to solutionising and aging, and correlate this response with detailed microstructural analysis.

## 4. Conclusions

Selective laser melted (SLMed) AA7075 in three different conditions, viz. as-SLMed, solutionised + quenched, and solutionised + quenched and aged at 120°C for 24 h, was explicitly studied in the context of microstructure, hardness, electrochemical response and localised corrosion morphology, and compared with wrought AA7075-T6. The following conclusion were drawn from the present study:

1. With similar bulk composition, the SLM technique generated a different and unique microstructure, with varying second phase particle type, size, shape and distribution as compared to that of wrought AA7075-T6. The as-SLMed AA7075 contained a quasicrystalline phase (ν-phase) as the principal nanostructural feature.

2. The microstructure of as-SLMed AA7075 varied with solutionising, and following ageing at 120°C for 24 h. Quasicrystalline ν-phase and melt pools were dissolved following solutionising. Ageing resulted in η′-phase ($MgZn_2$) precipitate formation within grains and along grain boundaries, revealing columnar grains in the alloy microstructure.

3. The hardness of as-SLMed AA7075 decreased with solutionising, as a result of quasicrystalline ν-phase dissolution. Artificial ageing following solutionising increased hardness; and following ~24 h ageing, the hardness was similar to that of the as-SLMed condition, as a result of η′-phase precipitate formation.

4. The electrochemical response as determined by the polarisation test conditions herein revealed passive-like behaviour for SLMed AA7075 in 0.1 M NaCl, in contrast to wrought AA7075-T6 which is noted for freely dissolving at open circuit conditions. Impedance spectroscopy validated the assertion that SLMed AA7075 presents greater corrosion resistance than wrought AA7075-T6.

5. Localised corrosion following immersion in 0.1 M NaCl for 10h was principally associated with the dissolution of Al-matrix at the periphery of second phase particles in SLMed AA7075 and wrought AA7075-T6. However, pits generated upon SLMed AA7075 were found to be notably smaller when compared to those upon wrought AA7075-T6 – owing to finer microstructural features and the absence of large constituent particles. The pit distribution, size and number density of SLMed AA7075 varied with thermal treatment as a result of the corresponding microstructural variation. Pitting was more pronounced along melt pool boundaries in the as-SLMed AA7075 condition. Following solutionising, pits were generated uniformly in the Al-matrix. Ageing resulted in pits with size greater along grain boundaries than within grains.

**Table 1.** Composition (wt.%) of AA7075 powder, as-SLMed AA7075 and wrought AA7075-T6 as determined by ICP-OES; and nominal composition of wrought AA7075.

|  | Zn | Mg | Cu | Cr | Fe | Si | Al |
|---|---|---|---|---|---|---|---|
| **AA7075 powder** | 4.7 | 2.13 | 1.3 | 0.2 | 0.13 | 0.09 | Bal. |
| **as-SLMed AA7075** | 3.62 | 1.85 | 1.31 | 0.2 | 0.13 | 0.1 | Bal. |
| **Wrought AA7075-T6** | 5.55 | 2.39 | 1.47 | 0.19 | 0.27 | 0.06 | Bal. |
| **Nominal composition of wrought AA7075 [56]*** | 5.1-6.1 | 2.1-2.9 | 1.2-2 | 0.18-0.28 | 0.5 Max | 0.4 Max | Bal. |

**\****other elements may include:* **Ti** (0.2 Max)**, Zr** (0.05 Max)**, Mn** (0.3 Max)**.**

**Table 2.** Composition (at.%) of phases that were observed in SLMed AA7075, as qualitatively determined by STEM-EDXS.

| Specimen | Probable phase | Zn | Mg | Cu | Fe | Si | Cr | O | Al |
|---|---|---|---|---|---|---|---|---|---|
| as-SLMed | Mg-Cu-Zn(-Al) (ν-phase) | 8.3 | 15.9 | 12.5 | 0.5 | 0.2 | <0.1 | 1.8 | 60.8 |
| | Al-Cu-Fe-O | 0.7 | 0.9 | 21.9 | 10.0 | 1.0 | <0.1 | 18.4 | 47.1 |
| | Mg$_2$Si | 2.6 | 32.7 | 3.0 | 0.2 | 14.4 | <0.1 | 1.4 | 45.7 |
| | Al-matrix | 2.3 | 2.9 | 0.8 | 0.2 | < 0.1 | 0.1 | 1.5 | 92.2 |
| | Precursor to η′-phase (MgZn$_2$) | colspan: Unable to be determined | | | | | | | |
| Solutionised + quenched | Al-Fe-Si-Cu-Cr | 0.7 | 0.6 | 2.0 | 11.5 | 5.4 | 2.3 | 1.2 | 76.3 |
| | Mg-O | 1.1 | 14.9 | 0.6 | 0.2 | 0.2 | <0.1 | 16.6 | 66.4 |
| | Al-matrix | 1.4 | 1.9 | 0.5 | 0.2 | <0.1 | 0.1 | 0.7 | 95.2 |
| Solutionised + quenched, and aged at 120°C for 24 h | Al-Fe-Si-Cu-Cr | 1.3 | 1.4 | 1.1 | 3.7 | 1.7 | 0.7 | 0.3 | 89.8 |
| | Mg-Fe-O | 1.3 | 3.7 | 0.8 | 1.8 | 0.5 | 0.3 | 1.7 | 89.9 |
| | Al-matrix | 1.6 | 1.6 | 0.6 | 0.2 | <0.1 | 0.1 | 0.4 | 95.5 |
| | η′-phase (MgZn$_2$) | colspan: Unable to be determined | | | | | | | |

**Table 3.** Typical results from the fitting of data from EIS testing of wrought and SLMed AA7075 in 0.1 M NaCl. Corresponding data and equivalent circuits are shown in Figure 7. Polarisation resistance ($R_p$) corresponds to the extrapolated intersection of the impedance data with $-Z'' = 0$ at very low frequencies (in Figure 7a).

| Specimen | $R_1$ ($\Omega.cm^2$) | $R_2$ ($\Omega.cm^2$) | $R_3$ ($\Omega.cm^2$) | $R_4$ ($\Omega.cm^2$) | $R_p$ ($\Omega.cm^2$) | $Q_1$ ($\mu F.s^{(a-1)}.cm^{-2}$) | $Q_2$ ($\mu F.s^{(a-1)}.cm^{-2}$) | $L_1$ ($\Omega.cm^2$) |
|---|---|---|---|---|---|---|---|---|
| Wrought AA7075-T6 | 87 | 3100 | 1200 | 3400 | 1899 | 14.2 | 462 | 9500 |
| as-SLMed AA7075 | 81 | 2560 | 17280 | n/a | 19840 | 18.6 | 358 | NA |
| SLMed AA7075 (solutionised + quenched) | 29 | 2108 | 14570 | n/a | 16678 | 29.7 | 745 | NA |
| SLMed AA7075 (solutionised + quenched, and aged at 120°C for 24 h) | 38 | 1112 | 5775 | n/a | 6886 | 37.3 | 1278 | NA |

**Table 4.** Statistics related to pits developed upon wrought and SLMed AA7075 specimen surfaces following exposure to 0.1 M NaCl for 10 h, as determined from SEM image analysis (and with the aid of ImageJ®).

| Specimen | Pit location | Average pit number over 1000 μm² area | Pit size range (μm) | Average pit opening length or diameter (μm) |
|---|---|---|---|---|
| **Wrought AA7075-T6** | Grains and grain boundaries | 60 | 0.083 to 22 | 1.81 |
| **as-SLMed AA7075** | Melt pool boundaries | 430 | 0.085 to 2.58 | 0.63 ± 0.49 |
| | Melt pool core | 1510 | 0.063 to 0.52 | 0.23 ± 0.08 |
| **SLMed AA7075 (solutionised + quenched)** | At nanoscale particles | 790 | 0.13 to 1.46 | 0.39 ± 0.23 |
| | At sub-micrometre to micrometre sized particles | 20 | 0.41 to 6.83 | 2.19 ± 1.59 |
| **SLMed AA7075 (solutionised + quenched, and aged at 120°C for 24 h)** | Grains | 340 | 0.12 to 0.72 | 0.28 ± 0.12 |
| | Grain boundaries | 30 | 0.39 to 5.74 | 1.89 ± 1.1 |

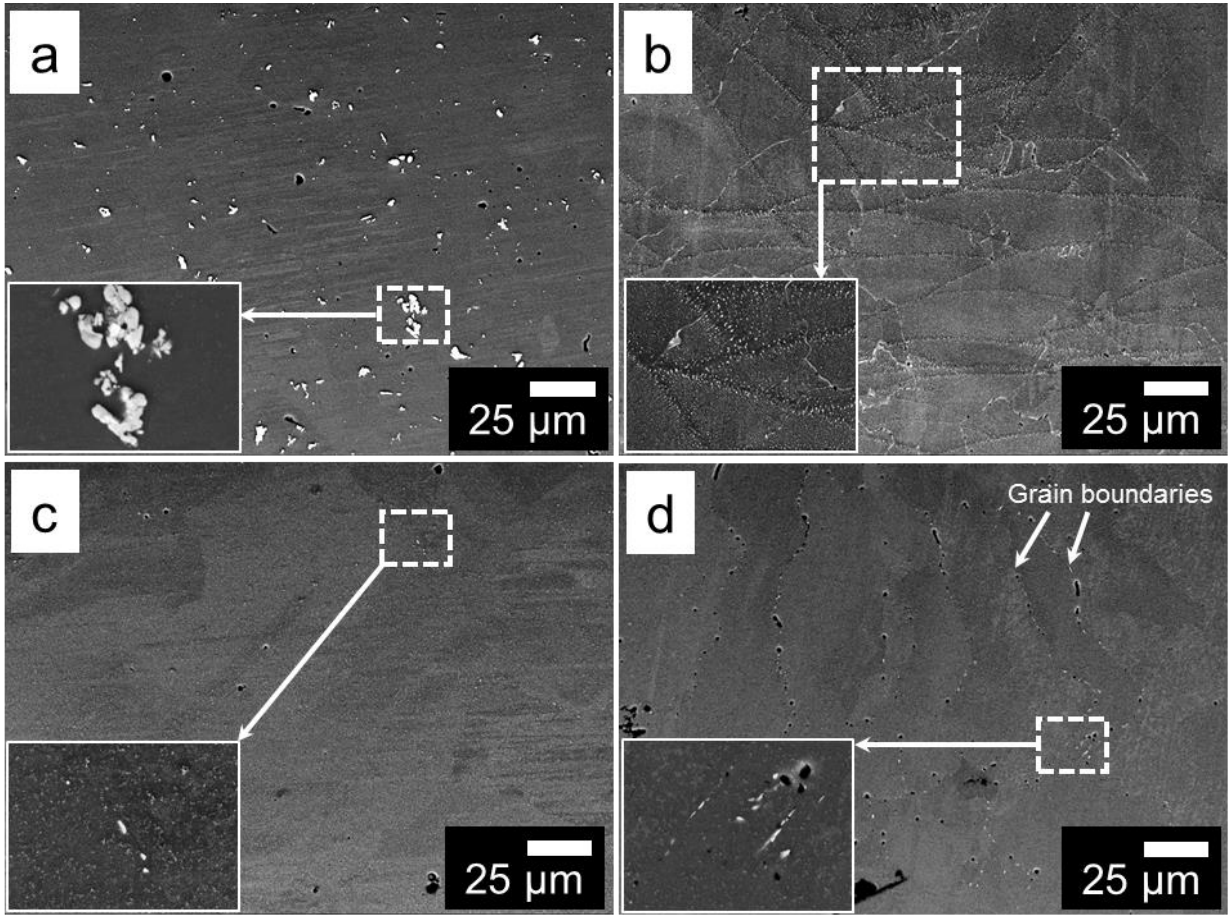

**Figure 1.** Backscattered electron images of AA7075 in the following conditions: (a) wrought T6, (b) as-SLMed, (c) SLMed (solutionised + quenched) and (d) SLMed (solutionised + quenched, aged at 120°C for 24 h). The inset, at bottom left, in each image corresponds to the high magnification image of the region within the box.

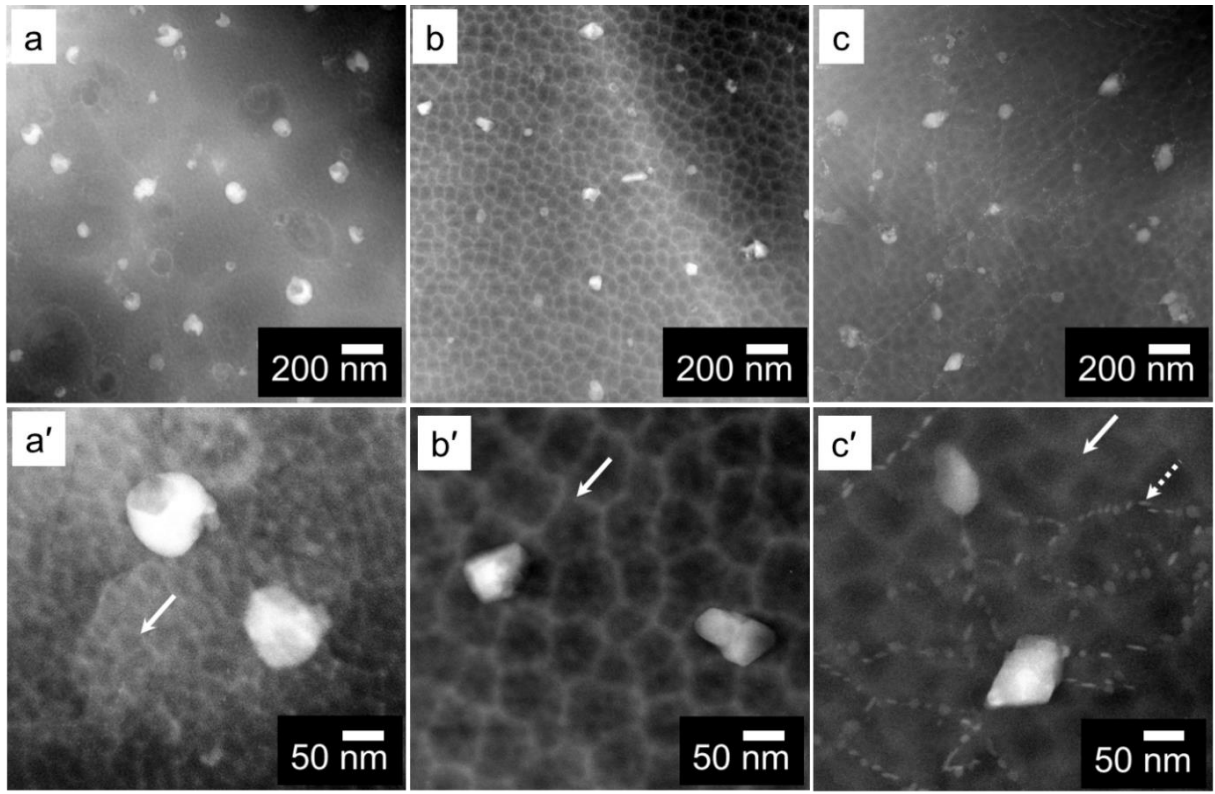

**Figure 2.** Low and high magnification HAADF-STEM images of SLMed AA7075 in the following conditions: (a and a′) as-SLMed, (b and b′) solutionised + quenched, and (c and c′) solutionised + quenched, and aged at 120°C for 24 h. Solid arrows point to cellular sub-structures. The dotted arrow in (c′) point to nanometre scale precipitates.

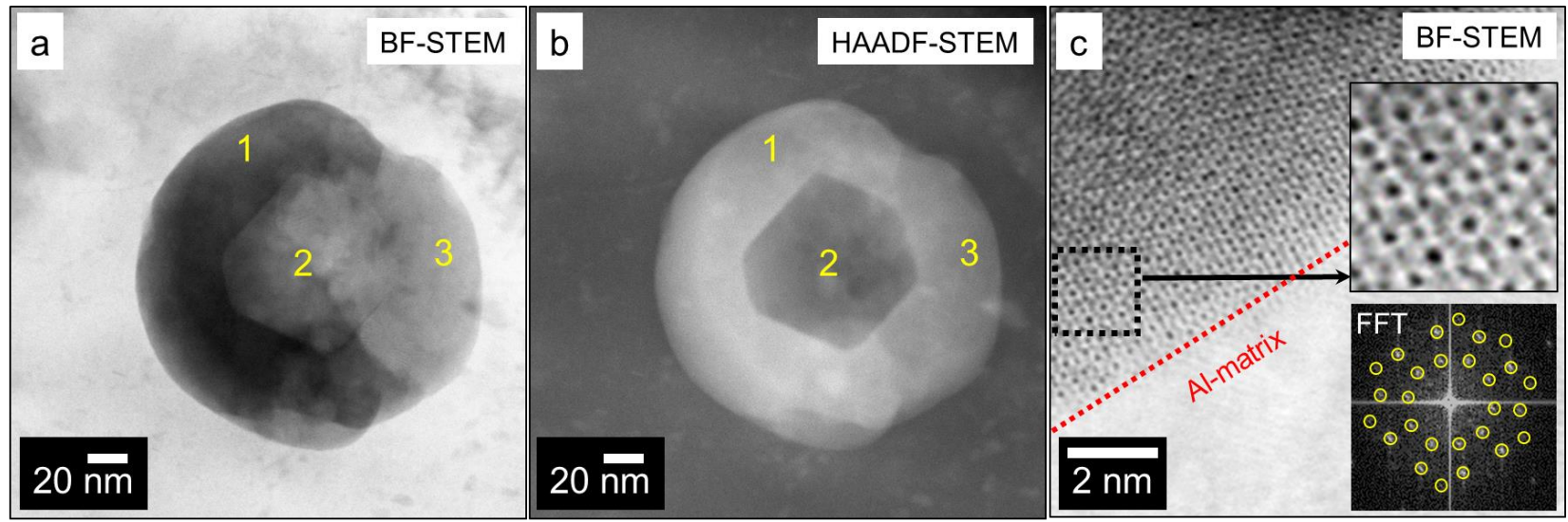

**Figure 3**. (a) BF-STEM and (b) HAADF-STEM images of three different particles, which are indicated by numbers, co-nucleated in the Al-matrix of as-SLMed AA7075. (c) The atomic resolution BF-STEM image of the particle region annotated as '1' in (a) and (b) reveals an icosahedral quasicrystal structure with 5-fold symmetry. The inset in (c) with circles, corresponds to the diffraction pattern from the quasicrystalline phase and was generated from the fast-Fourier Transform (FFT) of the corresponding image. The dotted line in (c) is the boundary between the quasicrystalline particle 1 and the Al-matrix. Particles 1 and 2 were characterised as the quasicrystalline ν-phase containing Mg-Zn-Cu-(Al) and particle 2 was determined to be $Mg_2Si$.

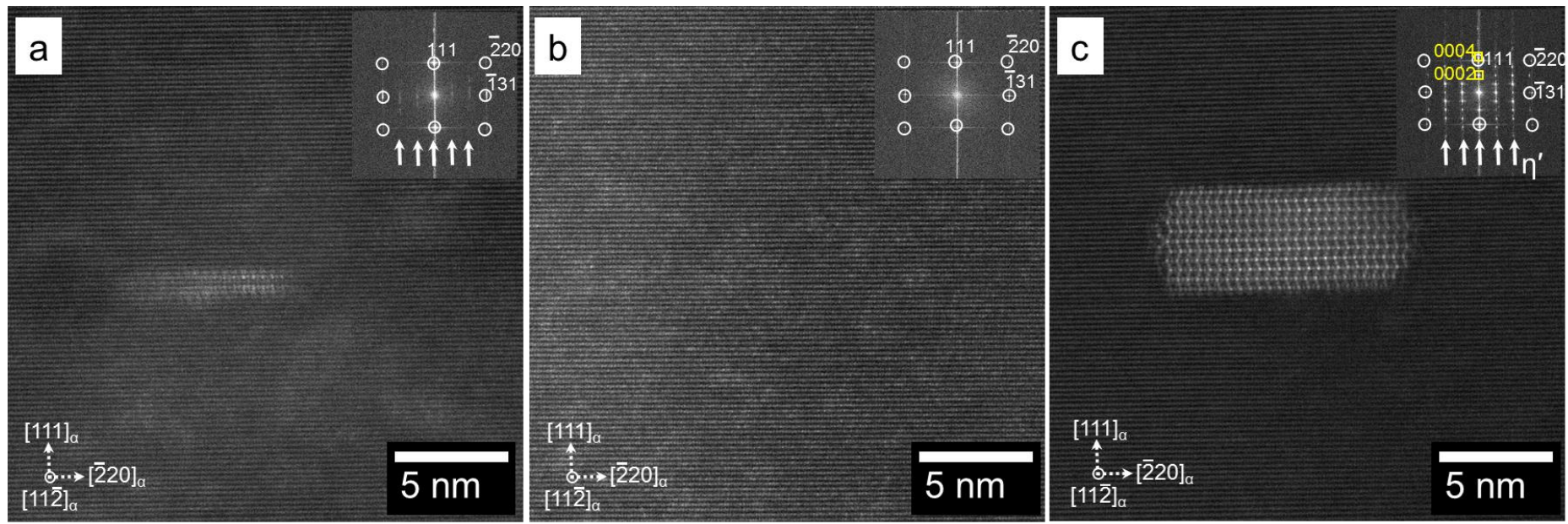

**Figure 4.** Atomic resolution HAADF-STEM images of SLMed AA7075 in the following conditions: (a) as-SLMed, (b) solutionised + quenched and (c) solutionised + quenched, and aged at 120°C for 24 h. Insets in images (a-c) are diffraction patterns obtained by the fast Fourier Transformation of the corresponding image. Arrows on the inset in (a) point to the diffraction from the precursor of η′-phase precipitate, whilst in (c) point to diffraction from η′-phase. Circles on the insets in (a), (b) and (c) correspond to the diffraction from the Al-matrix, whilst square boxes in the inset in (c) correspond to diffraction from 0002 and 0004 planes of η′-phase.

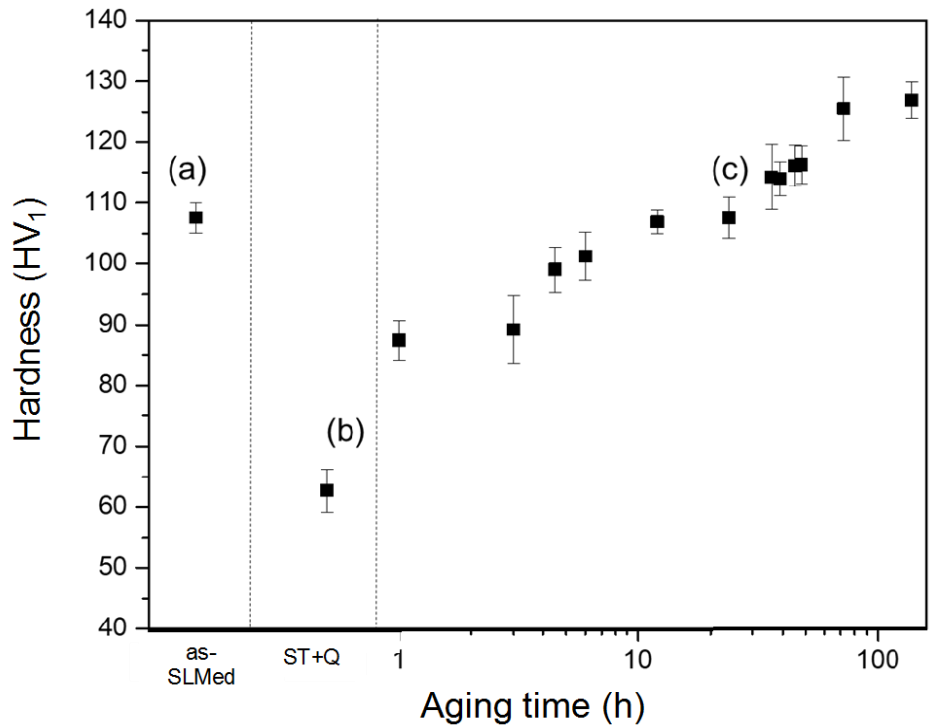

**Figure 5.** Hardness evolution of SLMed AA7075 in the following conditions: (a) as-SLMed, (b) solutionised + quenched (ST+Q), and solutionised + quenched, and aged at 120°C (where (c) denotes 24 h ageing). The average Vickers hardness number of wrought AA7075-T6 was determined to be 197 ± 5 VHN.

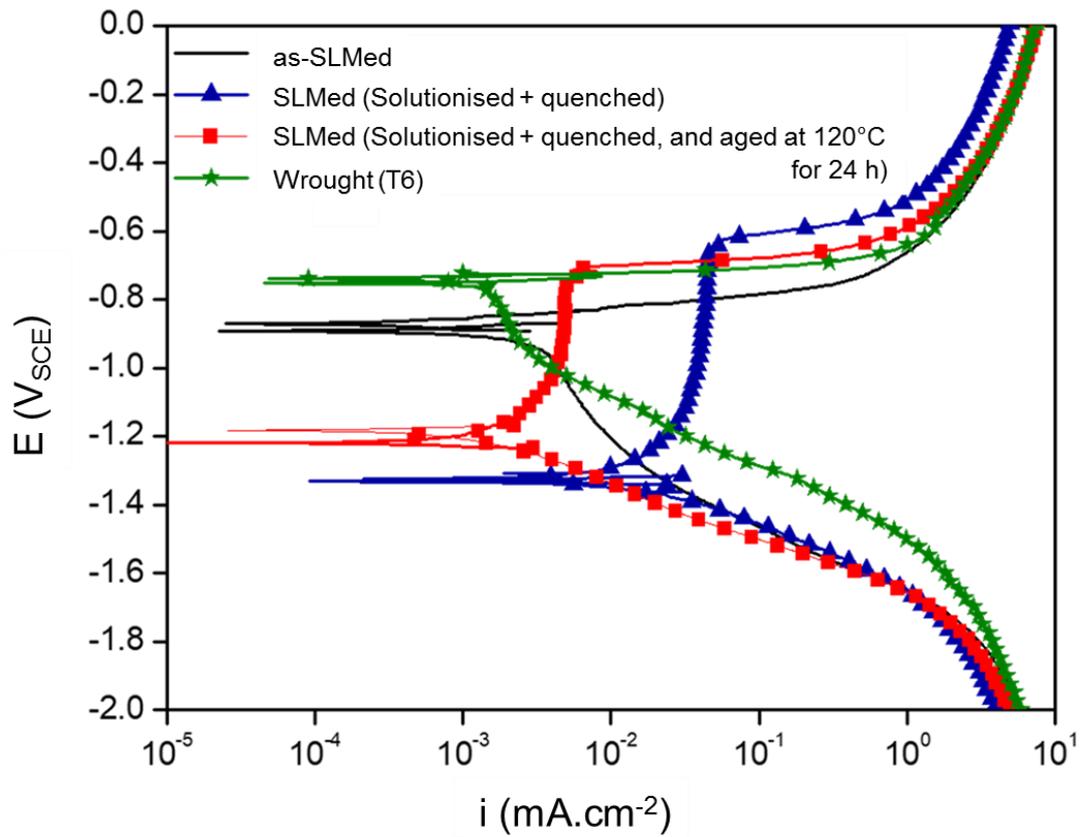

**Figure 6.** Potentiodynamic polarisation response of wrought and SLMed AA7075 in deaerated 0.1 M NaCl. The plots presented are comprised from a unique anodic and unique cathodic scan for each of the specimen conditions.

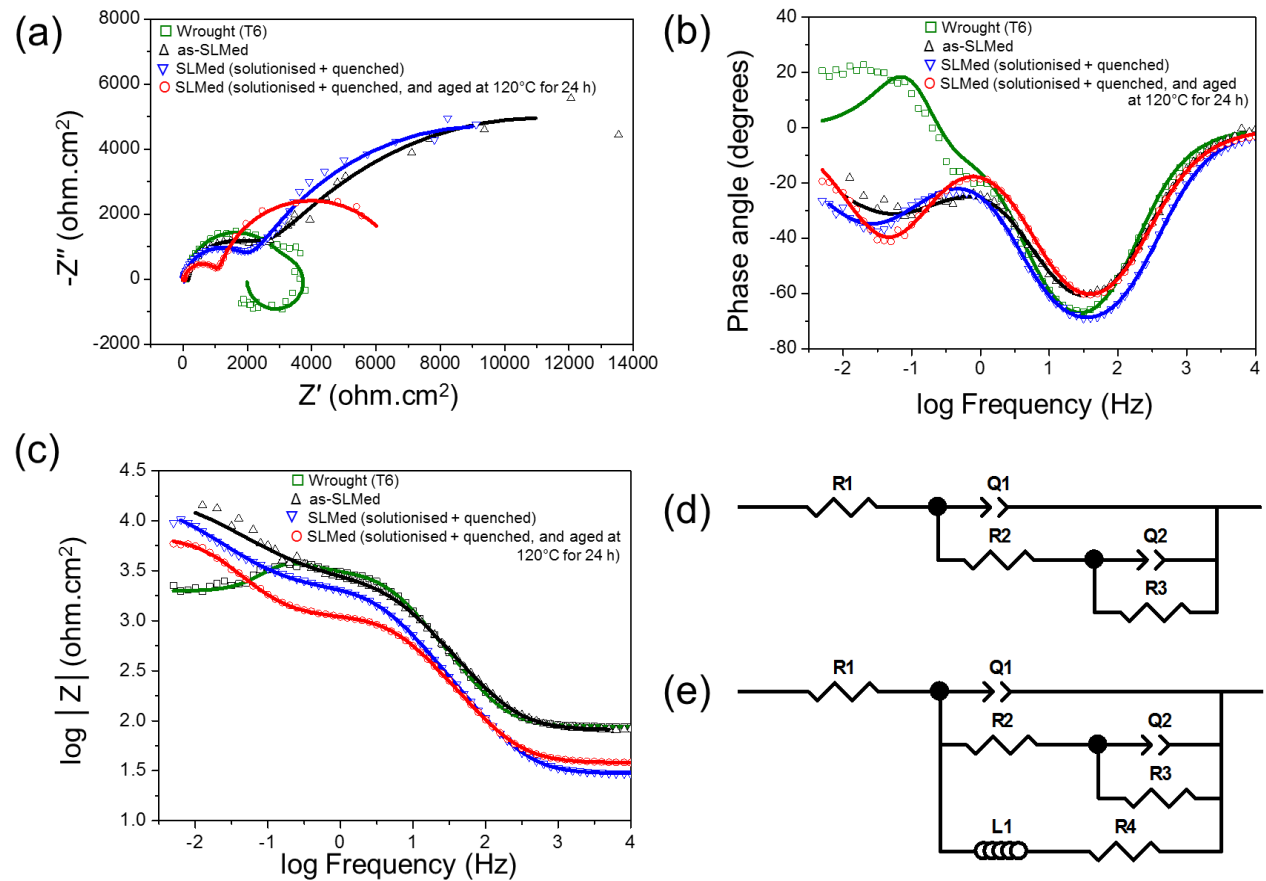

**Figure 7.** EIS data of wrought and SLMed AA7075 in 0.1 M NaCl: (a) Nyquist plots, (b) Bode Phase angle plost and (c) Bode magnitude plots. Experimental data are presented as data points, whilst solid lines are derived from the model fit. (d) and (e) are the equivalent circuits for SLMed and wrought AA7075, respectively.

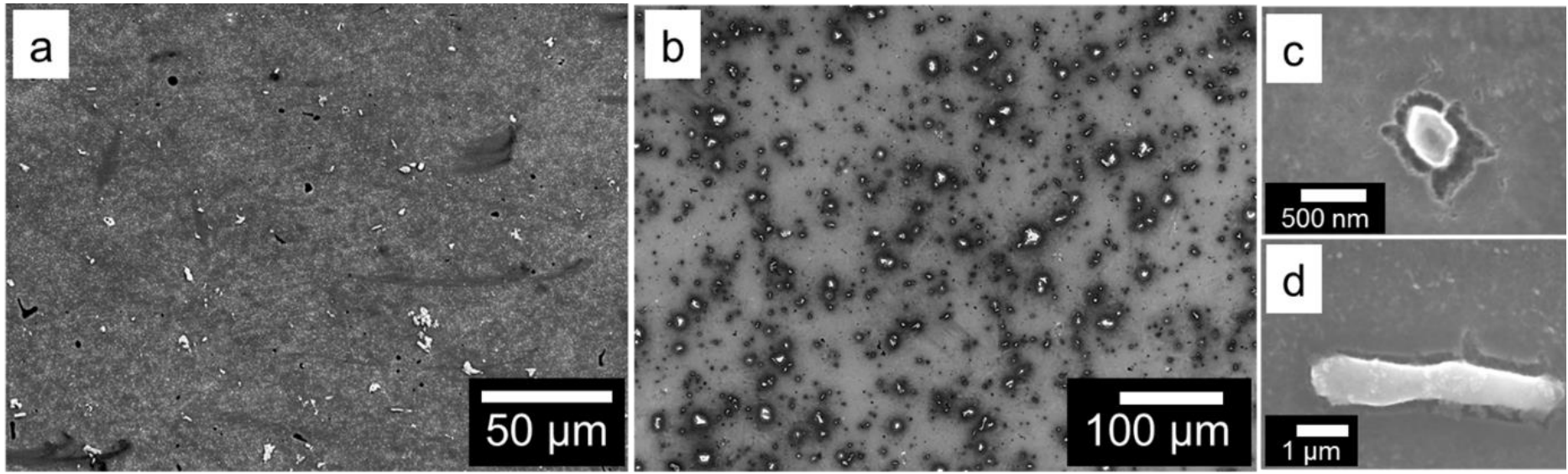

**Figure 8.** Secondary electron images of wrought AA7075-T6: (a) prior to exposure to 0.1 M NaCl and (b-d) following 10 h exposure to 0.1 M NaCl. High magnification images of constituent particles in image (b) are shown in images (c) and (d).

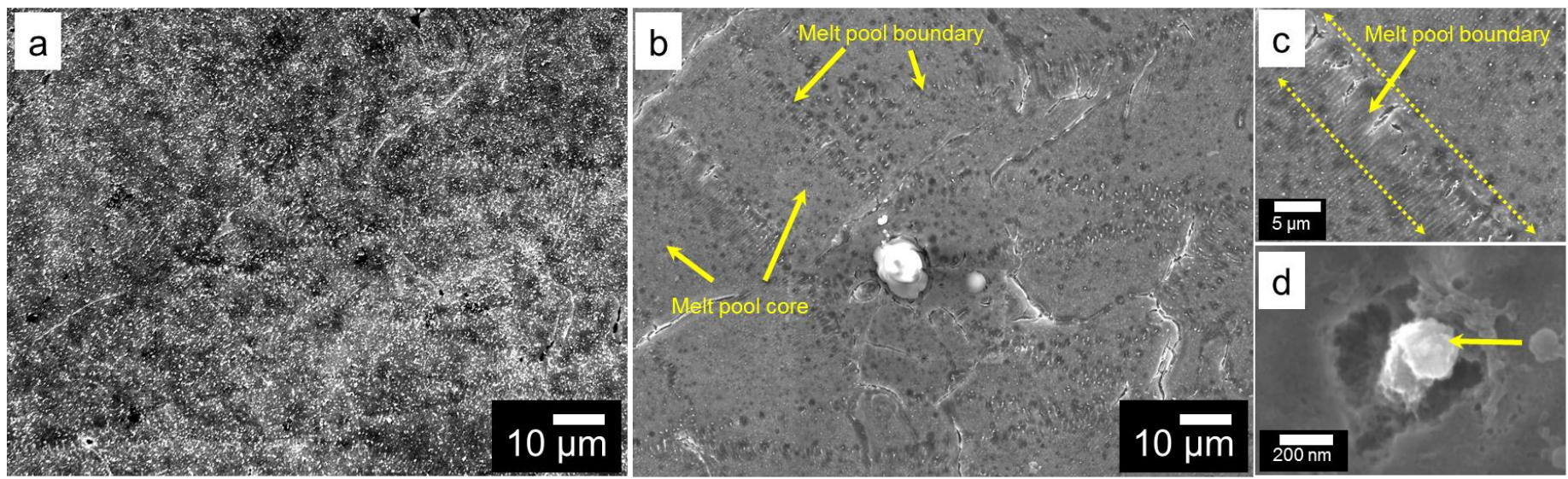

**Figure 9.** Secondary electron images of as-SLMed AA7075: (a) prior to exposure to 0.1 M NaCl and (b-d) following 10 h exposure to 0.1 M NaCl. Specimen surfaces in (a) and (b) are from the identical location. High magnification images of a melt pool boundary and a nanoscale particle form the core of melt pools in image (b) are shown in images (c) and (d), respectively.

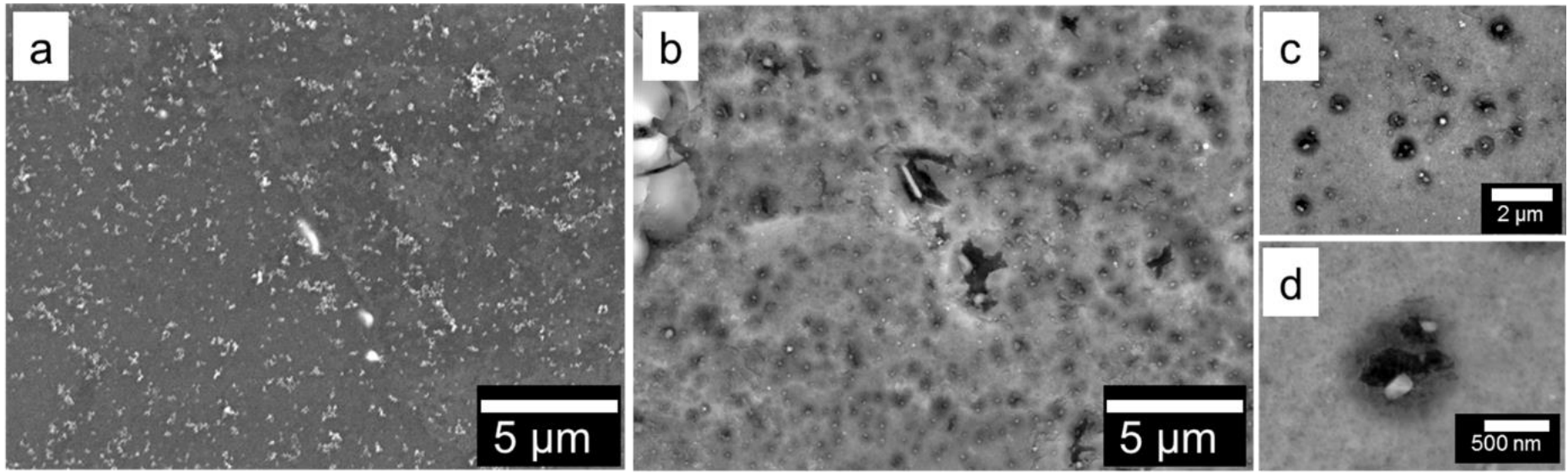

**Figure 10.** Secondary electron images of SLMed AA7075 in the solutionised + quenched condition: (a) prior to exposure to 0.1 M NaCl and (b-d) following 10 h exposure to 0.1 M NaCl. Specimen surfaces in (a) and (b) are from the identical location. High magnification images of nanoscale particles in image (b) are shown in images (c) and (d).

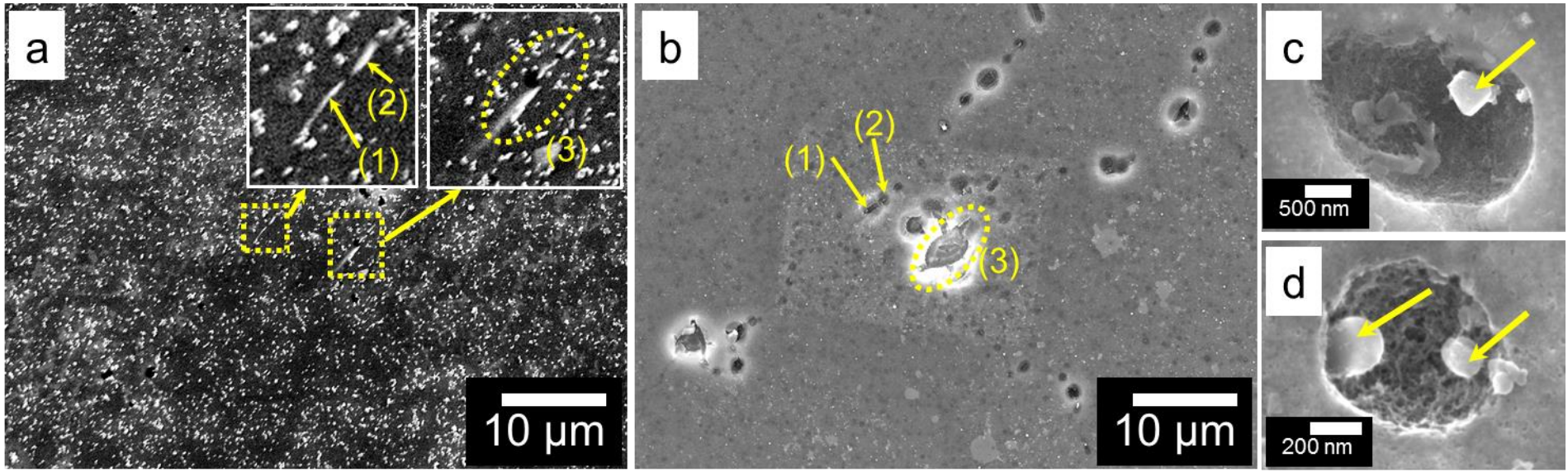

**Figure 11.** Secondary electron images of SLMed AA7075 in the solutionised + quenched, and aged at 120°C for 24 h condition: (a) prior to exposure to 0.1 M NaCl and (b-d) following 10 h exposure to 0.1 M NaCl. The insets in (a) are high magnification images of the regions within the dotted boxes. Sites (1), (2) and (3) in the insets in image (a), and in image (b) correspond to each other. High magnification images of nanoscale particles in image (b) are shown in images (c) and (d).